\documentclass[paper=letter, fontsize=12pt]{article}
\usepackage[english]{babel} % English language/hyphenation
\usepackage{amsmath,amsfonts,amsthm} % Math packages
\usepackage[utf8]{inputenc}
\usepackage{float}
\usepackage{lipsum} % Package to generate dummy text throughout this template
\usepackage{blindtext}
\usepackage{graphicx}
\usepackage{caption}
\usepackage{subcaption}
\usepackage[sc]{mathpazo} % Use the Palatino font
\usepackage[T1]{fontenc} % Use 8-bit encoding that has 256 glyphs
\usepackage{microtype} % Slightly tweak font spacing for aesthetics
\usepackage[hmarginratio=1:1,top=32mm,columnsep=20pt]{geometry} % Document margins
\usepackage{multicol} % Used for the two-column layout of the document
\usepackage{booktabs} % Horizontal rules in tables
\usepackage{float} % Required for tables and figures in the multi-column environment - they need to be placed in specific locations with the [H] (e.g. \begin{table}[H])
\usepackage{hyperref} % For hyperlinks in the PDF
\usepackage{bookmark}
\usepackage{lettrine} % The lettrine is the first enlarged letter at the beginning of the text
\usepackage{paralist} % Used for the compactitem environment which makes bullet points with less space between them
\usepackage{abstract} % Allows abstract customization
\usepackage{titlesec} % Allows customization of titles

\renewcommand\thesection{\Roman{section}} % Roman numerals for the sections
\renewcommand\thesubsection{\Roman{subsection}} % Roman numerals for subsections

\titleformat{\section}[block]{\large\scshape\centering}{\thesection.}{1em}{} % Change the look of the section titles
\titleformat{\subsection}[block]{\large}{\thesubsection.}{1em}{} % Change the look of the section titles
\usepackage{fancyhdr} % Headers and footers
\title{\vspace{-15mm}\fontsize{24pt}{10pt}\selectfont\textbf{A series representation for the Black-Scholes formula}} % Article title
\author{
{Jean-Philippe Aguilar }\\[2mm]
{\it BRED Banque Populaire, Modeling Department, 18 quai de la Râpée, Paris - 75012}\\[2mm]
{jean-philippe.aguilar@bred.fr}
}%\\ % Your name
\date{}

\providecommand{\keywords}[1]{\textbf{\textit{Key words---}} #1}
\newcommand{\res}{\mathrm{Res}}
\newcommand{\id}{\mathrm{d}}

\usepackage{mathtools}

\begin{document}
\maketitle % Insert title
\tableofcontents
\pagestyle{headings}
\setcounter{page}{1}
\pagenumbering{arabic}
 \begin{abstract}
\noindent We prove and test an efficient series representation for the European Black-Scholes call (formula \eqref{BS_series}), which generalizes and refines previously known approximations, and works in every market configuration.
\end{abstract}
% \keywords{Mellin transform, Multidimensional complex analysis, Fractional analysis, Stable distribution, Europen option pricing}


\thispagestyle{fancy} % All pages have headers and footers

\section{Introduction}

\noindent The celebrated Black-Scholes formula \cite{BS73}
\begin{equation}\label{BlackScholesFormula}
V(S,K,r,\sigma,\tau) \, = \, SN(d_+) - Ke^{-r\tau} N(d_-) \hspace*{1cm} d_{\pm} = \frac{1}{\sigma\sqrt{\tau}}\left(\log\frac{S}{K}+r\tau \right) \pm \frac{1}{2}\sigma\sqrt{\tau}
\end{equation}
where $N(.)$ is the normal distribution function, is a popular tool for the pricing of European options of strike K and time-to-maturity $\tau=T-t$, where the market conditions are described by an underlying (spot) asset price $S$, volatility $\sigma$ and risk-free interest rate $r$. 
\newline
\noindent It is an interesting question to know whether the Black-Scholes formula can be approximated by some simple terms. Under the strong assumption that the asset is "at-the-money forward", that is when
\begin{equation}
S \, \simeq \, Ke^{-r\tau}
\end{equation}
then there exists an approximation (see the paper by Brenner and Subrahmanyam \cite{BS94}, cited by \cite{Willmott06}):
\begin{equation}\label{Brenner}
V(S,K,r,\sigma,\tau) \, \simeq \, 0.4 \, S \,  \sigma\sqrt{\tau}
\end{equation}
This approximation can be useful in certain situations but has some major drawbacks:
\begin{enumerate}
\item[-] It is not very precise: for instance when $K=4000, r=1\%, \sigma=20\%, \tau = 1Y$ then the approximation \eqref{Brenner} yields
\begin{equation}
V(4000 \times e^{-0.01 \times 1} = 3960.2 , 4000, 0.01, 0.2, 1) \, \simeq \, 3960 \times 0.4 \times 0.2\sqrt{1} \, = \, 316.82
\end{equation}
while the usual Black-Scholes formula \eqref{BlackScholesFormula} yields
\begin{equation}
V(3960.2 , 4000, 0.01, 0.2, 1) \, = \, 315.45
\end{equation}
\item[-] Moreover, the "ATM forward" condition is very restrictive; one would prefer an approximation that works in any market condition, even if some extra terms should be added to \eqref{Brenner}.
\end{enumerate}

\noindent A natural idea is to Taylor expand the r.h.s. of \eqref{BlackScholesFormula}, for instance in powers of $\frac{S}{K}$. But, as already noticed by Estrella in \cite{Estrella95}, the Black-Scholes formula mixes two components of strongly different natures: the logarithmic function, possessing an expansion converging very fast but only for a small range of arguments, and the normal distribution, whose expansion (in odd powers of the arguments) converges on the whole real axis but with fewer level of accuracy. As shown by Estrella, this leads to situations where, for a plausible range of parameter values, the Taylor series for \eqref{BlackScholesFormula} diverges.

\noindent The main motivation of this note is therefore to derive a simple, efficient and compact approximation formula for the Black-Scholes formula \eqref{BlackScholesFormula}, that works in every market situation and can be made as more precise as one wishes; such a formula is obtained under the form of a fast converging series \eqref{BS_series}. This series, whose terms are very simple and straightforward to compute, converges as quickly as $\sigma\sqrt{\tau}$ is small, and, in any case, is absolutely convergent.
\newline 

\noindent The paper is organized as follows: first we derive the series formula for the pricing of the European call. The starting point is the Green representation for the solution of the Black-Scholes PDE; the proof uses tools from complex analysis in $\mathbb{C}^2$. We discuss our formula in the ATM-forward case, and show that is also constitutes a refinement of the Brenner-Subrahmanyam approximation in this case. Then we demonstrate the quickness of convergence of the series and show that the results are extremely close to the Black-Scholes formula after only a few iterations, in any market situation (out of, at or in the money). At the end of the paper, we make a short review of complex analysis in one and several dimensions, so that the reader who would be unfamiliar with these concepts can find them conveniently presented, and be provided with some classic references in the literature.

\section{Pricing formula}

The Black-Scholes model is a Gaussian model into which the underlying asset price $S$ is assumed to be described by a geometric Brownian motion of drift $r$ and volatility $\sigma$. From standard delta-hedging and non-arbitrage arguments, it can be shown that the price $V(S,K,r,\sigma,t)$ of an European call option of strike K and maturity $T$ satisfies the Black-Scholes equation, which is a partial differential equation (PDE) with terminal condition \cite{BS73}:
\begin{align}\label{BS_Equation}
\left\{
\begin{aligned}
 & \frac{\partial V}{\partial t} \, + \, \frac{1}{2} \sigma^2 S^2\frac{\partial^2 V}{\partial S^2} \, + \, r S \frac{\partial V}{\partial S} \, - \, rV \, = \, 0  \hspace*{1cm} t\in[0,T]  \\
 & V(S,K,r,\sigma,t=T) \, = \, [S-K]^+
\end{aligned}
\right.
\end{align}

\subsection{The Green function approach}
It is known (see the classic textbook by Wilmott \cite{Willmott06} for instance) that, with the change of variables
\begin{align}
\left\{
\begin{aligned}
 & x \, := \, \log S \, + \, (r-\frac{\sigma^2}{2}) \, \tau \\
 & \tau \, := \, T-t \\
 & V(S,K,r,\sigma,t)  \, := \, e^{-r\tau}W(x,K,r,\sigma,\tau)
\end{aligned}
\right.
\end{align}
then the Black-Scholes PDE \eqref{BS_Equation} resumes to the diffusion (or heat) equation
\begin{equation}
\frac{\partial W}{\partial \tau} \, - \, \frac{\sigma^2}{2}\frac{\partial^2 W}{\partial x^2} \, = \, 0
\end{equation}
whose fundamental solution (i.e., Green function) is the heat kernel:
\begin{equation}
g(x,K,r,\sigma,\tau) \, := \, \frac{1}{\sigma\sqrt{2\pi\tau}} \, e^{-\frac{x^2}{2\sigma^2\tau}}
\end{equation}
In this new set of variables the terminal condition becomes an initial condition:
\begin{equation}
W(x,K,r,\sigma,\tau = 0) \, = \, [e^x-K]^+
\end{equation}
and therefore, by the method of Green functions, we know that we can express $W$ as
\begin{equation}
W(x,K,r,\sigma,\tau) \, = \, \int\limits_{-\infty}^{+\infty} \, [e^{x+y}-K]^{+} \, g(y,K,r,\sigma,\tau) \, \id y
\end{equation}
Turning back to the initial variables (we keep the notation for time to maturity $\tau$):
\begin{equation}\label{BS_GreenForm}
V(S,K,r,\sigma,\tau) \, = \, e^{-r\tau} \, \int\limits_{-\infty}^{+\infty} [Se^{(r-\frac{\sigma^2}{2})\tau+y}-K]^{+} \, \frac{1}{\sigma\sqrt{2\pi\tau}}e^{-\frac{y^2}{2\sigma^2 \tau}} \, \id y
\end{equation}

\subsection{The call price as a complex integral}
Let us introduce the notations
\begin{equation}
z \, := \, \sigma\sqrt{\tau} \,\,\,\, , \,\,\,\, [\log]:= \log\frac{S}{K} + r\tau
\end{equation}
Then we can write:
\begin{equation}
[Se^{(r-\frac{\sigma^2}{2})\tau+y}-K]^{+} \, = \,  K \, [ e^{[\log] - \frac{z^2}{2} +y} -1 ]^{+}
\end{equation}
and rewrite the price \eqref{BS_GreenForm} under the form:
\begin{align}\label{BS_2}
V(S,K,r,\sigma,\tau) & = \frac{Ke^{-r\tau}}{\sqrt{2\pi}} \,  \int\limits_{\frac{z^2}{2}-[\log]}^{\infty} \, (e^{[\log] - \frac{z^2}{2} +y} -1 ) \, \frac{1}{z} e^{-\frac{y^2}{2z^2}} \id y \, 
\end{align}

\noindent
Let us introduce a Mellin-Barnes representation for the heat kernel-term in \eqref{BS_2} (see \eqref{Cahen} in Appendix, and \cite{Flajolet95,Erdélyi54} or any monograph on integral transforms):
\begin{equation}
\frac{1}{z} e^{-\frac{y^2}{2z^2}} \, = \, \frac{1}{z} \, \int\limits_{c_{1}-i\infty}^{c_{1}+i\infty} \Gamma(t_1)  \left( \frac{y^2}{2z^2}  \right)^{-t_1} \, \frac{\id t_1}{2i\pi} \hspace*{1cm} ( c_{1}> 0 )
\end{equation}
We thus have:
\begin{equation}
V(S,K,r,\sigma,\tau) \, = \, \frac{Ke^{-r\tau}}{\sqrt{2\pi}} \, \int\limits_{c_{1}-i\infty}^{c_{1}+i\infty} 2^{t_1} \Gamma(t_1) \, \int\limits_{\frac{z^2}{2}-[\log]}^{\infty} \, (e^{[\log] - \frac{z^2}{2} +y} -1 ) \,y^{-2 t_1} \, \id y \, z^{2t_1-1} \, \frac{\id t_1}{2i\pi} 
\end{equation}
Integrating by parts in the $y$-integral yields:
\begin{equation}
V(S,K,r,\sigma,\tau) \, = \, \frac{Ke^{-r\tau}}{\sqrt{2\pi}} \, \int\limits_{c_{1}-i\infty}^{c_{1}+i\infty} 2^{t_1} \frac{\Gamma(t_1)}{2t_1-1} \, \int\limits_{\frac{z^2}{2}-[\log]}^{\infty} \, e^{[\log] - \frac{z^2}{2} +y} \,y^{1-2 t_1} \, \id y \, z^{2t_1-1} \, \frac{\id t_1}{2i\pi} 
\end{equation}

\noindent
Let us introduce another Mellin-Barnes representation for the remaining exponential term (again, see \eqref{Cahen} in Appendix):
\begin{equation}
e^{[\log]-\frac{y^2}{2}+y} \, = \, \int\limits_{c_{2}-i\infty}^{c_{2}+i\infty} \, (-1)^{-t_2} \, \Gamma(t_2)  \left( [\log]-\frac{z^2}{2}+y   \right)^{-t_2} \, \frac{\id t_2}{2i\pi} \hspace*{1cm} ( c_{2}> 0 )
\end{equation}
Therefore the call price is:
\begin{multline}\label{BS_3}
V(S,K,r,\sigma,\tau) \, = \, \frac{Ke^{-r\tau}}{\sqrt{2\pi}} \times \\
\int\limits_{c_{1}-i\infty}^{c_{1}+i\infty} \int\limits_{c_{2}-i\infty}^{c_{2}+i\infty}  \, (-1)^{-t_2} \, \frac{2^{t_1}}{2t_1-1} \, \Gamma(t_1) \Gamma(t_2) \, \int\limits_{\frac{z^2}{2}-[\log]}^{\infty} y^{1-2t_1} \left( [\log]-\frac{z^2}{2}+y   \right)^{-t_2} \id y \, z^{2t_1-1} \, \frac{\id t_1}{2i\pi} \wedge \frac{\id t_2}{2i\pi}
\end{multline}
The $y$-integral is a particular case of Bêta-integral \cite{Abramowitz72} and equals:
\begin{equation}
\int\limits_{\frac{z^2}{2}-[\log]}^{\infty} y^{1-2t_1} \left( [\log]-\frac{z^2}{2}+y   \right)^{-t_2} \id y \, = \, \left( \frac{z^2}{2} - [\log]  \right)^{2-2t_1-t_2} \frac{\Gamma(1-t_2)\Gamma(-2+2t_1+t_2)}{\Gamma(2t_1-1)}
\end{equation}
and converges on the conditions $Re(t_2)<1$ and $Re(2t_1+t_2)>2$; plugging into \eqref{BS_3}, using the Gamma function functional relation $(2t_1-1)\Gamma(2t_1-1)=\Gamma(2t_1)$ and the Legendre duplication formula for the Gamma function \cite{Abramowitz72}:
\begin{equation}
\frac{\Gamma(t_1)}{\Gamma(2t_1)} \, = \, 2^{1-2t_1} \sqrt{\pi} \, \frac{1}{\Gamma(t_1+\frac{1}{2})}
\end{equation}
we obtain:
\begin{multline}\label{BS_4}
V(S,K,r,\sigma,\tau) \, = \, Ke^{-r\tau} \times \\
\int\limits_{c_{1}-i\infty}^{c_{1}+i\infty} \int\limits_{c_{2}-i\infty}^{c_{2}+i\infty}  \, (-1)^{-t_2} \, 2^{\frac{1}{2}-t_1} \, \frac{\Gamma(t_2)\Gamma(1-t_2)\Gamma(-2+2t_1+t_2)}{\Gamma(t_1+\frac{1}{2})} \left( \frac{z^2}{2} - [\log]  \right)^{2-2t_1-t_2} z^{2t_1-1}  \, \frac{\id t_1}{2i\pi} \wedge \frac{\id t_2}{2i\pi}
\end{multline}

\subsection{Residue summation}

Let us introduce the notations
\begin{equation}
\underline{t} \, := \, 
\begin{bmatrix}
t_1 \\ t_2
\end{bmatrix}
\hspace*{1cm}
\underline{c} \, := \, 
\begin{bmatrix}
c_{1} \\ c_{2} 
\end{bmatrix}
\hspace*{1cm}
\id \underline{t} \, := \, \id t_1 \wedge \id t_2
\end{equation}
and the \textit{complex differential 2-form}
\begin{equation}\label{omega}
\omega \, = \,  (-1)^{-t_2} \, 2^{\frac{1}{2}-t_1} \, \frac{\Gamma(t_2)\Gamma(1-t_2)\Gamma(-2+2t_1+t_2)}{\Gamma(t_1+\frac{1}{2})} \left( \frac{z^2}{2} - [\log]  \right)^{2-2t_1-t_2} z^{2t_1-1} \, \frac{\id \underline{t}}{(2i\pi)^2}
\end{equation}
then we can write the call price under the form:
\begin{equation}\label{V_form}
V(S,K,r,\sigma,\tau) \, = \, Ke^{-r\tau} \, \int\limits_{\underline{c} + i\mathbb{R}^2} \, \omega
\end{equation}

\noindent
where $\underline{c}$ is located in the polyhedra of convergence of the double Mellin-Barne integral \eqref{BS_4}
\begin{equation}
P \, = \, \{ \underline{t}\in\mathbb{C}^2 , \, Re(2t_1+t_2)>2, Re(t_2)>0 , Re(t_2)<1   \}
\end{equation}

\noindent
This complex integral can be performed by means of summation of $\mathbb{C}^2$-residues, by virtue of a multidimensional analogue to the residue theorem valid for this specific class of integrals (see \eqref{Tsikh_residue} and references in Appendix). This configuration is shown on fig. \ref{fig4}. Indeed, the characteristic quantity (see definition \eqref{Delta_n} in appendix) associated to the differential form \eqref{omega} is:
 
\begin{figure}[t]
\centering
\includegraphics[scale=0.5]{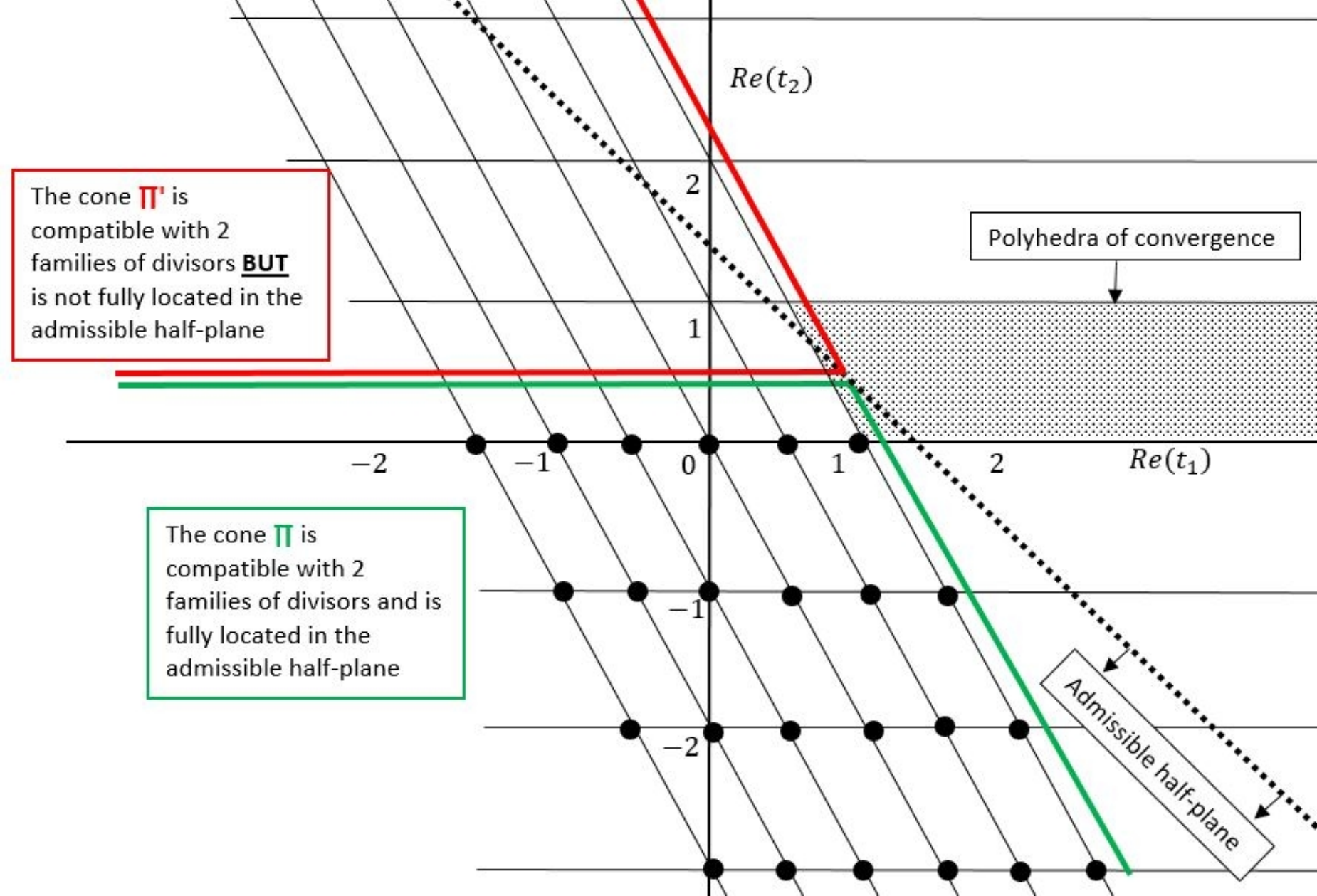}
\caption{The divisors $D_1$ (oblique lines) are induced by the $\Gamma(-2+2t_1+t_2)$ term, and $D_2$ (horizontal lines) by the $\Gamma(t_2)$ term. The intersection set $D_1 \cap D_2$ (the dots), located in the compatible green cone $\Pi$, gives birth to residues whose sum in the whole cone equals the integral \eqref{V_form}.}
\label{fig4}
\end{figure}

\noindent 
\begin{equation}
\Delta \, = \, 
\begin{bmatrix}
2 - 1 \\ 1 - 1 + 1
\end{bmatrix}
\, = \, 
\begin{bmatrix}
1 \\ 1
\end{bmatrix}
\end{equation}
and the admissible half-plane is 
\begin{equation}
\Pi_{\Delta} \, := \, 
\left\{
\underline{t}\in\mathbb{C}^2, \, Re (  \Delta \, . \, \underline{t}  ) \, < \,  \Delta . \underline{c} 
\right\}
\end{equation}
in the sense of the euclidean scalar product. This half-plane is therefore located under the line
\begin{equation}
t_2 \, = \, -t_1 \, + \, c_{1} \, + \, c_{2}
\end{equation}
In this half-plane, the cone $\Pi$ as shown of fig. \ref{fig4} and defined by 
\begin{equation}
 \Pi \, : = \, \left\{ \underline{t}\in\mathbb{C}^2 \, , \, Re(t_2) \leq 0 \, , \, Re(2t_1+t_2) \leq 2       \right\}
\end{equation}
contains and is compatible with the two family of divisors
\begin{equation}
\left\{
\begin{aligned}
& D_1 \, = \, \left\{\underline{t}\in\mathbb{C}^2, -2+2t_1 + t_2 = -n_1 \,\, , \,\,\, n_1 \in\mathbb{N} \right\}
\\
& D_2 \, = \, \left\{\underline{t}\in\mathbb{C}^2, t_2 = -n_2 \,\, , \,\,\, n_2 \in\mathbb{N} \right\}
\end{aligned}
\right.
\end{equation}
induced by $\Gamma(-2+2t_1+t_2)$ and $\Gamma(t_2)$ respectively. To compute the residues associated to every element of the singular set $ D_1 \cap D_2$, we change the variables:
\begin{equation}
\left\{
\begin{aligned}
& u_1 \, := \, -2 + 2t_1 + t_2 \\
& u_2 \, := \, t_2
\end{aligned}
\right.
\longrightarrow
\left\{
\begin{aligned}
& t_1 \, = \, \frac{1}{2} (2 + u_1 - u_2) \\
& t_2 \, = \, u_2 \\ 
& dt_1 \wedge dt_2 \, = \, \frac{1}{2} \, du_1 \wedge du_2
\end{aligned}
\right.
\end{equation}
so that in this new configuration $\omega$ reads
\begin{equation}
\omega \, = \, \frac{1}{2} (-1)^{-u_2} \, 2^{\frac{u_2-u_1-1}{2}} \, \frac{\Gamma(u_2)\Gamma(1-u_2)\Gamma(u_1)}{\Gamma(1+\frac{u_1-u_2+1}{2})} \left( \frac{z^2}{2} - [\log]  \right)^{-u_1} z^{u_1-u_2+1} \, \frac{\id u_1}{2i\pi } \wedge \frac{\id u_2}{2i\pi}
\end{equation}
With this new variables, the divisors $D_1$ and $D_2$ are induced by the $\Gamma(u_1)$ and $\Gamma(u_2)$ functions in $(u_1,u_2)=(-n,-m)$, $n,m\in\mathbb{N}$. From the singular behavior of the Gamma function around a singularity \eqref{sing_Gamma}, we can write
\begin{multline}
\omega \, \underset{(u_1,u_2)\rightarrow (-n,-m)}{\sim} \, \\
\frac{1}{2} (-1)^{-u_2} \, \frac{(-1)^{n+m}}{n!m!} \, 2^{\frac{u_2-u_1-1}{2}} \, \frac{\Gamma(1-u_2)}{\Gamma(1+\frac{u_1-u_2+1}{2})} \left( \frac{z^2}{2} - [\log]  \right)^{-u_1} z^{u_1-u_2+1} \, \frac{\id u_1}{2i\pi (u_1+n)} \wedge \frac{\id u_2}{2i\pi (u_2+m)}
\end{multline}
and therefore the residues are, by the Cauchy formula:
\begin{equation}
\res(u_1=-n , u_2=-m) \, = \,  (-1)^{n} \, 2^{\frac{n-m-1}{2}-1} \, \frac{1}{n!\Gamma(1+\frac{m-n+1}{2})}  \,  \left( \frac{z^2}{2} - [\log]  \right)^{n} z^{m-n+1}
\end{equation}
By virtue of the residue theorem \eqref{Tsikh_residue}, the sum of the residues in the whole cone equals the integral \eqref{V_form}:
\begin{equation}\label{V1-series}
V(S,K,r,\sigma,\tau) \, = \, Ke^{-r\tau} \sum\limits_{n,m=0}^{\infty} \, (-1)^{n} \, 2^{\frac{n-m-1}{2}-1} \, \frac{1}{n!\Gamma(1+\frac{m-n+1}{2})}  \,  \left( \frac{z^2}{2} - [\log]  \right)^{n} z^{m-n+1}
\end{equation}

\noindent
We can further simplify by changing the index $m\rightarrow m+1$ and introducing $Z:=\frac{z}{\sqrt{2}}= \frac{\sigma\sqrt{\tau}}{\sqrt{2}}$, and we finally obtain:
\begin{center}
\fbox{
\begin{minipage}[b]{\textwidth}
Let $[\log]:= \log\frac{S}{K} \, + \, r\tau$ and $Z:= \frac{\sigma\sqrt{\tau}}{\sqrt{2}} $, then the Black-Scholes price of the European call is given by the absolutely convergent double series:
\begin{equation} \label{BS_series}
V(S,K,r,\sigma,\tau) \, = \, \frac{Ke^{-r\tau}}{2} \sum\limits_{\substack{n = 0 \\ m = 1}}^{\infty} \, \frac{(-1)^n}{n!\Gamma(1+\frac{m-n}{2})}  \,  \left( Z^2 - [\log]  \right)^{n} Z^{m-n}
\end{equation}
\end{minipage}
}
\end{center}

\subsection{Brenner-Subrahmanyam and refinement}
By defintion of $[\log]$, the ATM forward configuration implies:
\begin{equation}
S \, = \, Ke^{-r\tau} \, \Longrightarrow \, [\log] \, = \, 0
\end{equation}
In this case, the series \eqref{BS_series} becomes
\begin{equation} \label{Brenner_series}
V(S,K,r,\sigma,\tau) \, = \, \frac{S}{2} \sum\limits_{\substack{n = 0 \\ m = 1}}^{\infty} \, \frac{(-1)^n}{n!\Gamma(1+\frac{m-n}{2})}  \, Z^{n+m}
\end{equation}
This series is now a series of positive powers of $Z$. It starts for $n=0, m=1$ and goes as follows:
\begin{equation}
V(S,K,r,\sigma,\tau) \, = \, \frac{S}{2}\frac{1}{\Gamma(\frac{3}{2})} \, Z \, + O(Z^2)  
\end{equation}
Recalling that $Z=\frac{\sigma\sqrt{\tau}}{\sqrt{2}}$ and that $\Gamma(\frac{3}{2}) = \frac{\sqrt{\pi}}{2}$ \cite{Abramowitz72}, we get
\begin{equation}
V(S,K,r,\sigma,\tau) \, = \, \frac{S}{\sqrt{2\pi}} \, \sigma\sqrt{\tau} \, + O(\sigma^2 \tau)  
\end{equation}
As:
\begin{equation}
\frac{1}{\sqrt{2\pi}} \, \simeq \, 0.399 \, \simeq \, 0.4
\end{equation}
we thus recover the Brenner-Subrahmanyam approximation \eqref{Brenner}:
\begin{equation}
V(S,K,r,\sigma,\tau) \, = \, 0.4 \, S \,  \sigma\sqrt{\tau} \, + O(\sigma^2 \tau)  
\end{equation}
Note that the series \eqref{Brenner_series} is indeed a refinement of the Brenner-Subrahmanyam approximation. In this precise market situation, the call price can therefore be expressed as a power series of $\sigma\sqrt{\tau}$, which is not the case in the general case as shown in \eqref{BS_series}, where negative powers of $\sigma\sqrt{\tau}$ and powers of $[\log]$ arise.

\section{Numerical tests}

\subsection{Quickness of convergence}

\begin{figure}[h]
 \centering
 \includegraphics[scale=0.3]{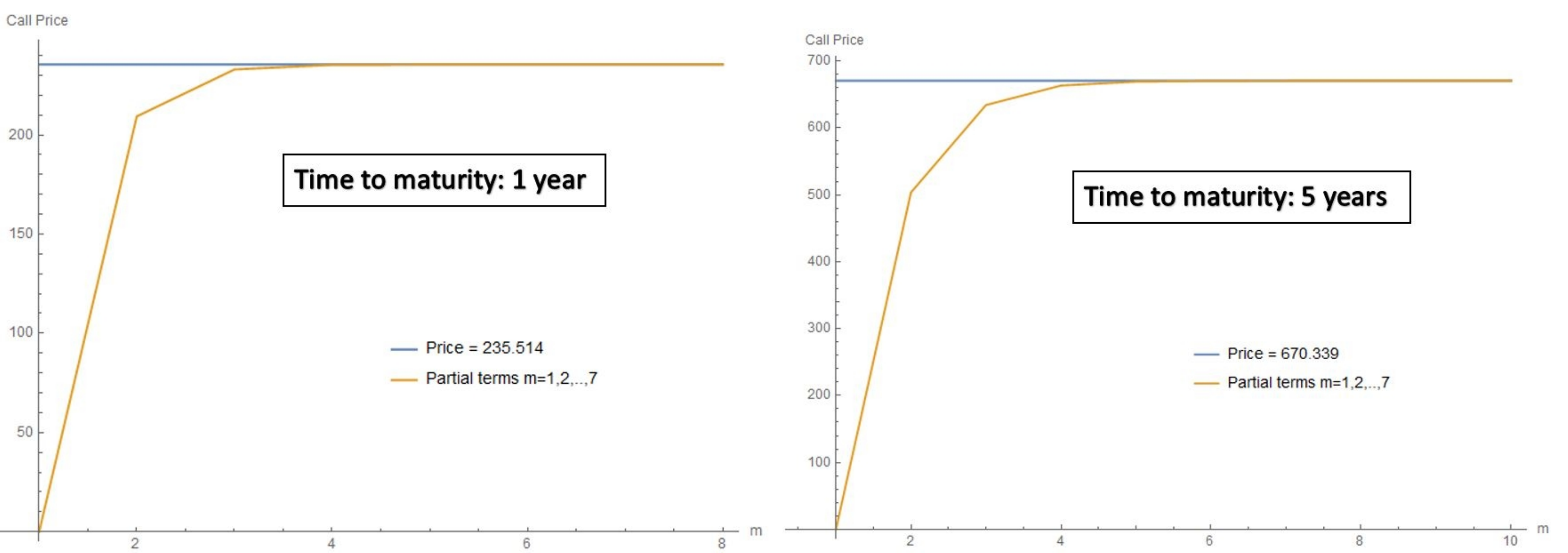}
\caption{Convergence of partial terms of the series \eqref{BS_series} to the call price, for short and long times to maturity (1Y and 5Y).}
 \label{fig5}
 \end{figure}

It is interesting to note that in the series \eqref{BS_series}, only the very first few terms contribute and already give an excellent approximation to the option price. See for instance Tab. \ref{fig:series}, where the set of parameters used is $S=3800, K=4000, r=1\%, \sigma= 20\%, \tau = 1$. The convergence is a little bit less fast when $\tau$ grows, but remains very efficient, see for instance Table \ref{fig:series5Y}. In Fig. \ref{fig5} we compute the partial terms of the series \eqref{BS_series} as a series of $m$, that is, the sum of vertical columns in tables \ref{fig:series} and \ref{fig:series5Y}, for a time to maturity $\tau$ being 1 or 5 years. We observe that, in both cases, the convergence is fast.

\begin{table}[h!]
\centering
\begin{scriptsize}
\begin{tabular}{|c||ccccccc|}
  \hline
  % after \\: \hline or \cline{col1-col2} \cline{col3-col4} ...
 {\bfseries $\tau$=1Y}  & 1 & 2 & 3 & 4 & 5 & 6 & 7   \\
  \hline
  \hline
  0 & 315.978 & 39.602 & 4.213 & 0.396 & 0.034 & 0.003 & 0.000   \\
  1 & -121.367 & -19.367 & -2.427 & -0.258 & -0.024 & -0.002 & -0.000   \\
  2 & 14.839 & 3.720 & 0.594 & 0.074 & 0.008  &  0.001 & 0.000   \\
  3 & 0 & -0.303 & -0.076 & -0.012 & -0.002 & -0.000 &  -0.000  \\
  4 & -0.116 & 0 & 0.005 & 0.001 & 0.000 & 0.000 & 0.000   \\
  5 & 0 & 0.001 & 0 & -0.000 & -0.000 & -0.000 & -0.000  \\
  6 & 0.001 & 0 & -0.000 & 0 & 0.000 & 0.000 & 0.000   \\
  \hline
  Call & 209.335 & 232.987 & 235.295 & 235.496 & 235.512 & 235.514 & 235.514    \\
  \hline
\end{tabular}
\end{scriptsize}
\caption{Table containing the numerical values for the $(n,m)$-term in the series (\ref{BS_series}) for the option price ($S=3800, \, K=4000, \, r=1\%, \sigma=20\%, \, \tau=1Y$). The call price converges to a precision of $10^{-3}$ after summing only very few terms of the series.}
\label{fig:series}
\end{table}

\begin{table}[h!]
\centering
\begin{scriptsize}
\begin{tabular}{|c||ccccccccc|}
  \hline
  % after \\: \hline or \cline{col1-col2} \cline{col3-col4} ...
 {\bfseries $\tau$=5Y}  & 1 & 2 & 3 & 4 & 5 & 6 & 7 & 8 & 9  \\
  \hline
  \hline
  0 & 678.845 & 190.246 & 45.256 & 9.512 & 1.810 & 0.317 & 0.052 & 0.008 & 0.001   \\
  1 & -192.706 & -68.762 & -19.271 & -4.584 & -0.0964 & -0.183 & -0.032 & -0.005 & -0.001  \\
  2 & 17.413 & 9.760 & 3.483 & 0.976 & 0.232  & 0.049 & 0.009  & 0.002 & 0.000 \\
  3 & 0 & -0.588 & -0.330 & -0.118 & -0.033 & -0.008 &  -0.002 & -0.000 & -0.000 \\
  4 & -0.074 & 0 & 0.015 & 0.008 & 0.003 & 0.001 & 0.000 & 0.000 & 0.000  \\
  5 & 0 & 0.002 & 0 & -0.000 & -0.000 & -0.000 & -0.000 &  -0.000 & -0.000 \\
  6 & 0.000 & 0 & -0.000 & 0 & 0.000 & 0.000 & 0.000  & 0.000 & 0.000 \\
  \hline
  Call & 503.477 & 634.134 & 663.288 & 669.082 & 670.131 & 670.306 & 670.334  & 670.338 & 670.338  \\
  \hline
\end{tabular}
\end{scriptsize}
\caption{Same set of parameters as in Table \ref{fig:series}, but for a longer maturity $\tau$= 5 years. The convergence is slightly slower but remains very efficient.}
\label{fig:series5Y}
\end{table}

% \begin{figure}[t]
% \centering
% \includegraphics[scale=0.7]{BSFig3.pdf}
% \caption{As shown in the matrix, the $(n,m)$-term in our expansion \eqref{BS_series} goes rapidly to $0$ when $n$ and $m$ grow. Only the first few terms are needed to obtain a very precise price. We demonstrate it on 3 evaluations (out of, at and in the money), where we only make 10 iterations and get an extremely precise price.}
% \label{fig3}
% \end{figure}

\subsection{Comparisons between the series and the BS formula}

\noindent In the following array, we compare the results obtained via an application of the Black-Scholes formula and via the expansion \eqref{BS_series} truncated at $n=m=20$. We show that for every time to maturity and every market situation (in, at or out of the money), this truncation is sufficient to agree with the Black-Scholes formula at a level of precision of $10^{-6}$. Other parameters are $K=4000, r=1\%, \sigma = 20\%$.

\vspace*{0.5cm}
\noindent
\begin{center}
\begin{tabular}{|c|c|c|c|c|c|c|c|c|}
\hline
\textbf{Out of the money (S=3800)}& $\tau = 1Y$ & $\tau = 2Y$ & $\tau = 3Y$ &$\tau = 4Y$ &$\tau = 5Y$  \\
\hline
\hline
Black-Scholes Formula & 235.5135954 & 376.3907685 &  488.2564760 & 584.4538077 & 670.3385381   \\
\hline
Series \eqref{BS_series}, 20 (n,m)-iterations & 235.5135954 & 376.3907685 & 488.2564760 & 584.4538077 & 670.3385381  \\
\hline
\hline
\textbf{At the money (S=400)}& $\tau = 1Y$ & $\tau = 2Y$ & $\tau = 3Y$ &$\tau = 4Y$ &$\tau = 5Y$  \\
\hline
\hline
Black-Scholes Formula & 337.3327476 & 486.1060719 &  603.0304375 & 703.1314722 & 792.2680293
   \\
\hline
Series \eqref{BS_series}, 20 (n,m)-iterations & 337.3327476 & 486.1060719 & 603.0304375 & 703.1314722 &  792.2680293 \\
\hline
\hline
\textbf{In the money (S=4200)}& $\tau = 1Y$ & $\tau = 2Y$ & $\tau = 3Y$ &$\tau = 4Y$ &$\tau = 5Y$  \\
\hline
\hline
Black-Scholes Formula & 458.7930654 & 609.5901660 &  728.9304673 & 831.3409473 & 922.6298574  \\
\hline
Series \eqref{BS_series}, 20 (n,m)-iterations & 458.7930654 & 609.5901660 & 728.9304673 & 731.3409473 &  922.6298574 \\
\hline
\end{tabular}
\end{center}

\section{Concluding remarks}

\noindent To the author's knowledge, formula \eqref{BS_series} is new. It can be efficiently used in practice as an alternative of the Black-Scholes formula, for instance in an excel sheet or a Mathematica file. More interesting, the technique used for obtaining this result applies to a wider class of problems. Indeed, as soon as one is able to write the option price under the form of an integral over a Green variable $y$:
\begin{equation}
V \, = \, \int \, Payoff \,\, \times \,\, Green \, function \,\,  \id y
\end{equation}
and that one can write a suitable Mellin-Barnes representation for the Green function, then the technology applies. This is notably the case for the class of L\'evy processes (of which Black-Scholes is a special case), where the Green functions are all known under the form of Mellin-Barnes representations (see for instance \cite{Mainardi05}, \cite{KK16} and references therein). This work is currently in progress \cite{ACKK17} .

\newpage
\appendix
\section{APPENDIX: Mellin transforms and residues}

We briefly present here some of the concepts used in the paper. The theory of the one-dimensional Mellin transform is explained in full detail in \cite{Flajolet95}. An introduction to multidimensional complex analysis can be found in the classic textbook \cite{Griffiths78}, and applications to the specific case of Mellin-Barnes integrals is developped in \cite{Tsikh94,Tsikh97,Aguilar08}.

\subsection{One-dimensional Mellin transforms}

1. The Mellin transform of a locally continuous function $f$ defined on $\mathbb{R}^+$ is the function $f^*$ defined by
\begin{equation}\label{Mellin_def}
f^*(s) \, := \, \int\limits_0^\infty \, f(x) \, x^{s-1} \, \id x
\end{equation}
The region of convergence $\{ \alpha < Re (s) < \beta \}$ into which the integral \eqref{Mellin_def} converges is often called the fundamental strip of the transform, and sometimes denoted $ < \alpha , \beta  > $.
\newline
\noindent 2. The Mellin transform of the exponential function is, by definition, the Euler Gamma function:
\begin{equation}
\Gamma(s) \, = \, \int\limits_0^\infty \, e^{-x} \, x^{s-1} \, \id x
\end{equation}
with strip of convergence $\{ Re(s) > 0 \}$. Outside of this strip, it can be analytically continued, expect at every negative $s=-n$ integer where it admits the singular behavior
\begin{equation}\label{sing_Gamma}
\Gamma(s) \, \underset{s\rightarrow -n}{\sim} \, \frac{(-1)^n}{n!}\frac{1}{s+n} \hspace*{1cm} n\in\mathbb{N}
\end{equation}
\newline
\noindent 3. The inversion of the Mellin transform is performed via an integral along any vertical line in the strip of convergence:
\begin{equation}\label{inversion}
f(x) \, = \, \int\limits_{c-i\infty}^{c+i\infty} \, f^*(s) \, x^{-s} \, \frac{\id s}{2i\pi} \hspace*{1cm} c\in ( \alpha, \beta )
\end{equation}
and notably for the exponential function one gets the so-called \textit{Cahen-Mellin integral}:
\begin{equation}\label{Cahen}
e^{-x} \, = \, \int\limits_{c-i\infty}^{c+i\infty} \, \Gamma(s) \, x^{-s} \, \frac{\id s}{2i\pi} \hspace*{1cm} c>0
\end{equation}
\newline
\noindent 4. When $f^*(s)$ is a ratio of products of Gamma functions of linear arguments:
\begin{equation}
f^*(s) \, = \, \frac{\Gamma(a_1 s + b_1) \dots \Gamma(a_n s + b_n)}{\Gamma(c_1 s + d_1) \dots \Gamma(c_m s + d_m)}
\end{equation}
then one speaks of a \textit{Mellin-Barnes integral}, whose \textit{characteristic quantity} is defined to be
\begin{equation}
\Delta \, = \, \sum\limits_{k=1}^n \, a_k \, - \, \sum\limits_{j=1}^m \, c_j
\end{equation}
$\Delta$ governs the behavior of $f^*(s)$ when $|s|\rightarrow \infty$ and thus the possibility of computing \eqref{inversion} by summing the residues of the analytic continuation of $f^*(s)$ right or left of the convergence strip:
\begin{equation}
\left\{
\begin{aligned}
& \Delta < 0 \hspace*{1cm} f(x) \, = \, -\sum\limits_{Re(s_N) > \beta} \, \res_{S_N} f^*(s)x^{-s}  \\
& \Delta > 0 \hspace*{1cm} f(x) \, = \, \sum\limits_{Re(s_N) < \alpha} \, \res_{S_N} f^*(s)x^{-s}
\end{aligned}
\right.
\end{equation}
For instance, in the case of the Cahen-Mellin integral one has $\Delta = 1$ and therefore:
\begin{equation}
e^{-x} \, = \, \sum\limits_{Re(s_n)<0} \res_{s_n} \Gamma(s) \, x^{-s} \, = \, \sum\limits_{n=0}^{\infty} \, \frac{(-1)^n}{n!}x^n
\end{equation}
as expected from the usual Taylor series of the exponential function.

\subsection{Multidimensional Mellin transforms}

1. Let the $\underline{a}_k$, $\underline{c}_j$, be vectors in $\mathbb{C}^2$,and the $b_k$, $d_j$ be complex numbers. Let $\underline{t}:=\begin{bmatrix} t_1 \\ t_2 \end{bmatrix}$ and $\underline{c}:=\begin{bmatrix} c_1 \\ c_2 \end{bmatrix}$ in $\mathbb{C}^2$ and "." represent the euclidean scalar product. We speak of a Mellin-Barnes integral in $\mathbb{C}^2$ when one deals with an integral of the type
\begin{equation}
\int\limits_{\underline{c}+i\mathbb{R}^2} \, \omega
\end{equation}
where $\omega$ is a complex differential 2-form who reads
\begin{equation}
\omega \, = \, \frac{\Gamma(\underline{a}_1.\underline{t}_1 + b_1) \dots \Gamma(\underline{a}_n.\underline{t}_n + b_n)}{\Gamma(\underline{c}_1.\underline{t}_1 + d_1) \dots \Gamma(\underline{c}_m.\underline{t}_m + b_m)} \, x^{-t_1} \, y^{-t_2} \, \frac{\id t_1}{2i\pi} \wedge \frac{\id t_2}{2i\pi} \hspace*{1cm} \, x,y \in\mathbb{R}
\end{equation}
The singular sets induced by the singularities of the Gamma functions
\begin{equation}
D_k \, := \, \{ \underline{t}\in\mathbb{C}^2 \, , \, \underline{a}_k.\underline{t}_k + b_k = -n_k \, , \, n_k \in\mathbb{N}   \} \,\,\,\, \, k=0 \dots n
\end{equation}
are called the \textit{divisors} of $\omega$. The \textit{characteristic vector} of $\omega$ is defined to be
\begin{equation}\label{Delta_n}
\Delta \, = \, \sum\limits_{k=1}^n \underline{a}_k \, - \, \sum\limits_{j=1}^m \underline{c}_j
\end{equation}
and the \textit{admissible half-plane}:
\begin{equation}
\Pi_\Delta \, := \, \{ \underline{t}\in\mathbb{C}^2 \, , \, \Delta . \underline{t} \, < \, \Delta . \underline{c}  \}
\end{equation}
\newline
\noindent 2. Let the $\rho_k$ in $\mathbb{R}$, the $h_k:\mathbb{C}\rightarrow\mathbb{C}$ be linear aplications and $\Pi_k$ be a subset of $\mathbb{C}^2$ of the type
\begin{equation}\label{Pik}
\Pi_k \, := \, \{ \underline{t}\in\mathbb{C}^2, \, Re(h_k(t_k)) \, < \, \rho_k \}
\end{equation}
A \textit{cone} in $\mathbb{C}^2$ is a cartesian product
\begin{equation}
\Pi \, = \, \Pi_1 \times \Pi_2
\end{equation}
where $\Pi_1$ and $\Pi_2$ are of the type \eqref{Pik}. Its \textit{faces} $\varphi_k$ are
\begin{equation}
\varphi_k \, := \, \partial \Pi_k \hspace*{1cm} k=1,2
\end{equation}
and its \textit{distinguished boundary}, or \textit{vertex} is
\begin{equation}
\partial_0 \, \Pi \, := \, \varphi_1 \, \cap \, \varphi_2
\end{equation}
\newline
3. Let $1<n_0<n$. We group the divisors $D=\cup_{k=0}^n \, D_k$ of the complex differential form $\omega$ into two sub-families
\begin{equation}
D_1 \, := \, \cup_{k=1}^{n_0} \, D_k \,\,\, \,\,\, D_2 \, := \, \cup_{k=n_0+1}^{n} \, D_k  \hspace*{1cm}  D \, = \, D_1\cup D_2
\end{equation}
We say that a cone $\Pi\subset\mathbb{C}^2$ is \textit{compatible} with the divisors family $D$ if:
\begin{enumerate}
\item[-] \, Its distinguished boundary is $\underline{c}$;
\item[-] \, Every divisor $D_1$ and $D_2$ intersect at most one of his faces:
\begin{equation}
D_k \, \cap \, \varphi_k \, = \, \emptyset \hspace*{1cm} k=1,2
\end{equation}
\end{enumerate}

\noindent 4. Residue theorem for multidimensional Mellin-Barnes integral \cite{Tsikh94,Tsikh97}: If $\Delta \neq 0$ and if $\Pi\subset\Pi_\Delta$ is a compatible cone located into the admissible half-plane, then
\begin{equation}\label{Tsikh_residue}
\int\limits_{\underline{c}+i\mathbb{R}^2} \, \omega \, = \, \sum\limits_{\underline{t}\in\Pi\cap (D_1\cap D_2)} \res_{\underline{t}} \, \omega
\end{equation}
and the series converges absolutely. The residues are to be understood as the "natural" generalization of the Cauchy residue, that is:
\begin{equation}
\res_0 \, \left[ f(t_1,t_2) \, \frac{\id t_1}{2i\pi t_1^{n_1}} \wedge \frac{\id t_2}{2i\pi t_1^{n_2}}  \right] \, = \, \frac{1}{(n_1-1)!(n_2-1)!}\frac{\partial ^{n_1+n_2-2}}{\partial t_1^{n_1-1} \partial t_2^{n_2-1} } f(t_1,t_2) |_{t_1=t_2=0}
\end{equation}

%----------------------------------------------------------------------------------------
%\end{multicols}
\end{document}